\documentclass[Physsubmission, Phys]{SciPost}

\hypersetup{
    colorlinks,
    linkcolor={red!50!black},
    citecolor={blue!50!black},
    urlcolor={blue!80!black}
}

\usepackage[bitstream-charter]{mathdesign}
\DeclareSymbolFont{usualmathcal}{OMS}{cmsy}{m}{n}
\DeclareSymbolFontAlphabet{\mathcal}{usualmathcal}

\begin{document}

\begin{center}{\Large \textbf{
Transversity through Dihadron: a constrained fit\\
}}\end{center}

\begin{center}
A. Courtoy\textsuperscript{1$\star$}
\end{center}

\begin{center}
{\bf 1} Instituto de F\'isica, Universidad Nacional Aut\'onoma de M\'exico\\
Apartado Postal 20-364, 01000 Ciudad de M\'exico, Mexico
\\

* aurore@fisica.unam.mx
\end{center}

\begin{center}
\today
\end{center}


\definecolor{palegray}{gray}{0.95}
\begin{center}
\colorbox{palegray}{
  \begin{tabular}{rr}
  \begin{minipage}{0.1\textwidth}
    \includegraphics[width=22mm]{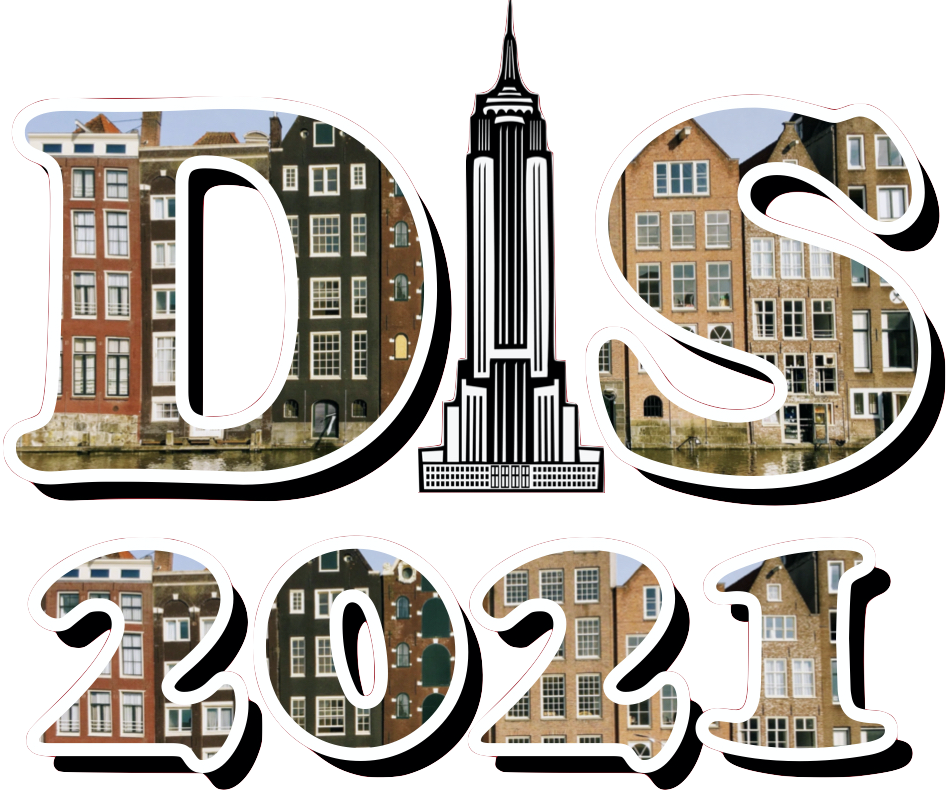}
  \end{minipage}
  &
  \begin{minipage}{0.75\textwidth}
    \begin{center}
    {\it Proceedings for the XXVIII International Workshop\\ on Deep-Inelastic Scattering and
Related Subjects,}\\
    {\it Stony Brook University, New York, USA, 12-16 April 2021} \\
    \doi{10.21468/SciPostPhysProc.?}\\
    \end{center}
  \end{minipage}
\end{tabular}
}
\end{center}

\section*{Abstract}
{\bf
We present a study of the valence transversity PDF obtained through a constrained fit. The effects of the constraints on the uncertainties of the PDF are explored. We show that the resulting isovector tensor charge is largely undetermined and, hence, compatible with all lattice evaluations. 
}

\vspace{10pt}
\noindent\rule{\textwidth}{1pt}
\tableofcontents\thispagestyle{fancy}
\noindent\rule{\textwidth}{1pt}
\vspace{10pt}

\section{Introduction}
\label{sec:intro}

Collinear Parton Distribution Functions (PDFs), at leading twist, reflect the distribution of quarks and gluons with a given spin configuration {\it w.r.t.} their parent hadron. Of the three leading-twist PDFs, the transversity PDF remains to be explored in wider kinematical regimes and complementary processes. The latter have been sparse due to the chiral-odd nature of the transversity PDF. While pioneering studies about transverse polarization of quarks focused on Drell-Yan processes~\cite{Artru:1989zv,Cortes:1991ja}, it is the advent of low-energy semi-inclusive DIS, together with the access to fragmentation functions in electron-positron annihilation processes, that provided the first phenomenological analyses~\cite{Anselmino:2007fs}. Those seminal works focused on transverse momentum dependent transversity; the analysis of dihadron fragmentation function~\cite{Jaffe:1997hf} allowed to remain in a fully collinear framework~\cite{Radici:2001na}.
Modern analyses, mainly based on single- and di-hadron SIDIS, have found statistically compatible valence transversity distributions, due to large uncertainties that would accommodate for tensions. In particular, pulls from the data for semi-inclusive DIS off deuteron  on the down distribution led  parametrizations to saturate the Soffer bounds. More recent global analyses~\cite{ Radici:2018iag,Benel:2019mcq,DAlesio:2020vtw,Cammarota:2020qcw,AbdulKhalek:2021gbh} have shown increasing trends towards: 1) the need to understand the role of positivity bounds on polarized PDFs -- the Soffer bound for the transversity PDF; 2) the interpretation of the sources of uncertainties. 
Contrarily to the other two leading-twist PDFs, the transversity is not constrained by sum rules. Rather, its first Mellin moment -- the tensor charge-- is a desired output of the (global) analyses, and its determination is affected by the various sources of uncertainties~\cite{Kovarik:2019xvh} to the  analysis of the transversity PDF.

In parallel, efforts to bridge low-energy QCD matrix elements to systematic searches for new fundamental interactions -- beyond the Standard Model-- have indicated a certain interest for form factors with small momentum transfer and charges beyond the $V-A$ structure~\cite{Bhattacharya:2011qm}. Pioneering studies~\cite{Courtoy:2015haa, Gao:2017ade} demonstrated the limitations on the then state-of-the-art determination of the tensor charge and the opportunities in combining efforts between lattice determinations and phenomenology. This supported the need for a deeper understanding of the first-principle constraints and their role on uncertainties.

In Ref.~\cite{Benel:2019mcq}, we took that path and provided a fit of the valence transversity PDFs with no explicit dependence on the Soffer bound. Nonetheless the latter was taken into account as a weight of the chi-square function and through constraints on the large-$x$ fall off.
In these proceedings, we aim to analyze further the uncertainty on the tensor charge that resulted from our constrained fit.

\section{Valence transversities}

The constrained fit proposed in Ref.~\cite{Benel:2019mcq} consists in an analysis of the dihadron SIDIS extraction of the transversity PDF including the positivity bounds as  constraints through the Lagrange multiplier method. This modality allowed for more flexibility in the parametrization to accommodate for the deuteron data~\cite{Adolph:2012nw}, which is sensitive to the down distribution. This is in contrast to imposing the positivity bound on the functional form~\cite{Anselmino:2007fs,Bacchetta:2012ty, Radici:2015mwa,Radici:2018iag}, for sketches on the differences see~\cite{Constantinou:2020hdm}.
The flexibility was further increased by weighting the chi-square by the probability of the Soffer bound to be fulfilled given the extracted proton and deuteron transversity combination from~\cite{Airapetian:2008sk,Adolph:2012nw}, taking hence into account existing tensions between the representation of the Soffer bound and the data.

We chose to express the functional form for the transversity in terms of Bernstein polynomials\footnote{The Bernstein polynomials are defined as
  \begin{eqnarray}
  B_{k, n}(x)&=& \left(\begin{array}{l}n\\ k \end{array}\right) x^k (1-x)^{n-k} \quad.
   \end{eqnarray}
 }, as has been done in Ref.~\cite{Dulat:2015mca},
 \begin{eqnarray}
x\,h_{1,i}^{q_v}\left(x ; p_{i,k}^q\right)&=&x^{1.25}\,\sum_{k=\{\kappa_{q,i}\}} \,p_{i,k}^q\, B_{k, n_i}\left(g(x)\right)\quad,
\label{eq:finalFF}
 \end{eqnarray}
 selecting the degrees of the polynomials in such a way as to flexibly span all kinematical regions. A rescaling of the variable will help the parametrization adjusting the data; we chose $g(x)=x^{0.3}$. Our resulting transversity PDFs feature a statistically representative error outside the data range, yet in agreement with the first principle constraints at hand. This has been achieved using four different parametrizations, {\it i.e.} four degrees for the polynomial $P_n$. A third step is taken to guide the down transversity to smoothly fall-off towards $0$ for $x\to 1$. The Lagrange multipliers method constrains $\left|x_l\,h_{1,i}^{d_v}\left(x_l ; p^{d\,II}_{i,k}\right)\right|< \epsilon_l$ for $x_l=\{0.3, 0.55\}$ and  $\epsilon_l=\{0.2, 0.1\}$, where $p^{d\,II}_{i,k}$ is the new optimal set of parameters. By constraining the large-$x$ behavior, our results qualitatively differ from the recent results of Ref.~\cite{Cammarota:2020qcw}.

\begin{figure}[h]
\centering
\includegraphics[width=0.495\textwidth]{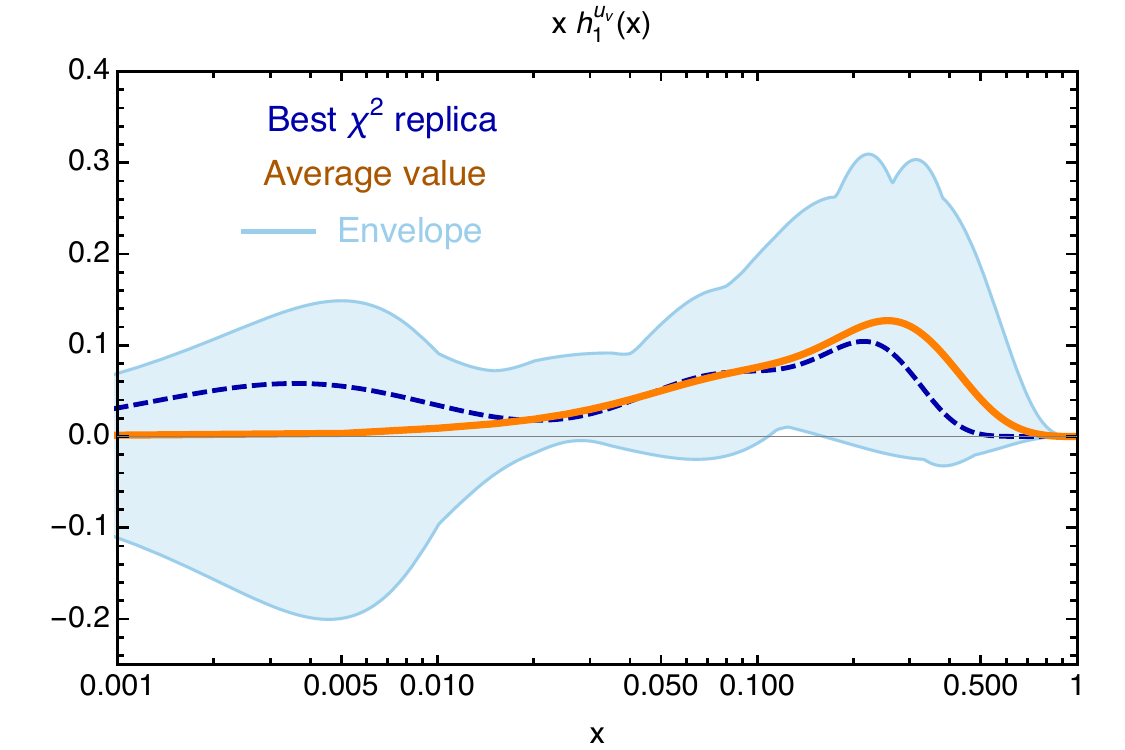}
\includegraphics[width=0.495\textwidth]{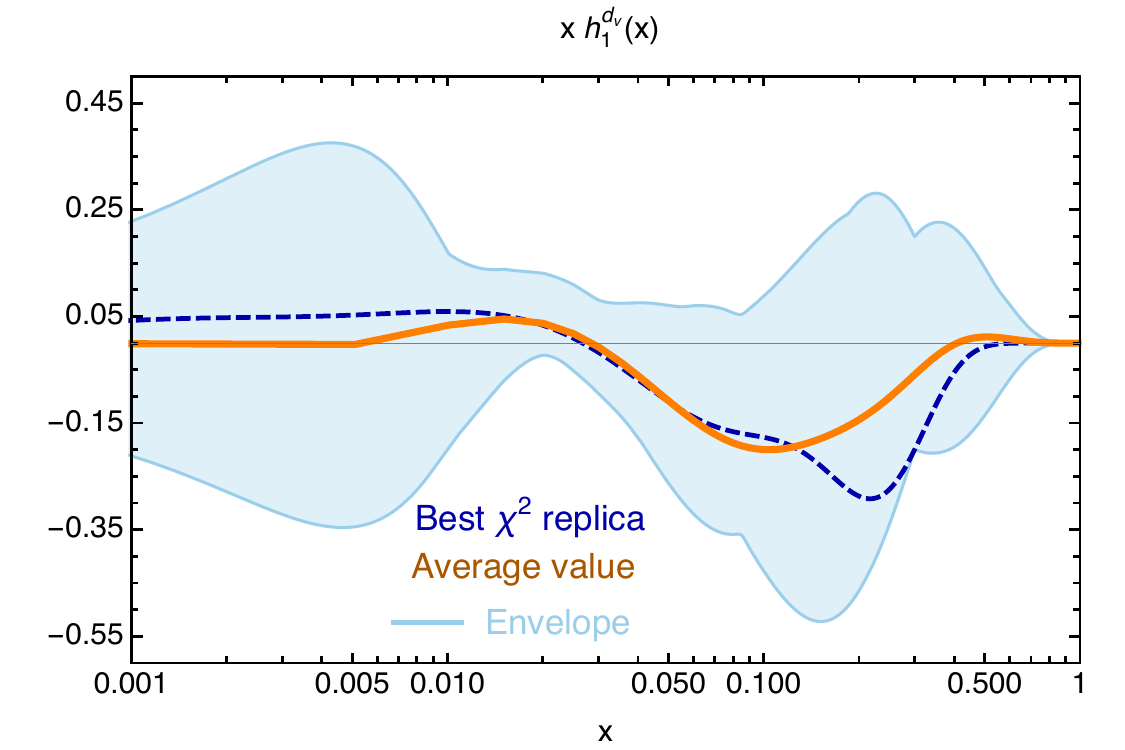}
\caption{Valence transversity PDFs: up (left), down(right). The light blue envelope represents the $N_{rep}$ fits, the orange curve the average of all $N_{rep}$ fits and the dashed blue curve the replica with best $\chi^2$ value. }
\label{fig:up_down}
\end{figure}

This 3-step procedure is repeated $N_{rep}=200$ times on the $N_{rep}$ sets of  proton and deuteron combinations randomly generated within their $1\sigma$ uncertainty. The resulting envelopes for the valence up and down transversity PDFs are shown in light blue in Fig.~\ref{fig:up_down}, respectively left and right. In the goal of studying the impact of the distribution of the $N_{rep}$ fits on the determination of the tensor charge, we compare the average over the $N_{rep}$ replicas (orange curve) to the replica with the best chi-square value, $\chi_{min}^2$, (blue dashed curve). The two curves almost coincide in the region covered by data, $x\in [0.0065,0.2871]$. However, their discrepancy at low and large values of $x$ reflects the need for more statistics in those regions. Studying the distribution of the replicas for a given value of $x$, we observe that the replicas are gaussianly distributed for low-$x$ values. This is no longer true for values of $x>x_{l=1}\sim 0.2871$.

\begin{figure}[h]
\centering
\includegraphics[width=0.65\textwidth]{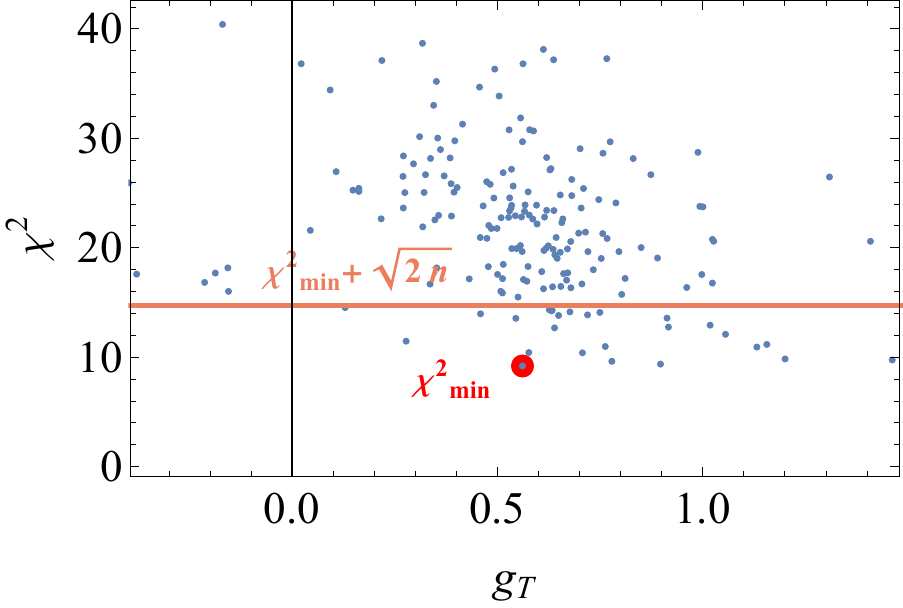}
\caption{ Distribution of the $N_{rep}=200$ values for $g_T$ plotted against $\chi^2$. The red dot corresponds to $(g_T(\chi_{min}^2),\chi_{min}^2)$. The orange line is set at $\chi_{min}^2+\sqrt{2 n}$, where $n$ is the number of degrees of freedom. }
\label{fig:gT_chi2}
\end{figure}
To understand the implication of this observation on the tensor charge, in Fig.~\ref{fig:gT_chi2}, we show the behavior of the $N_{rep}$ values for the tensor charge against the corresponding $\chi^2$. We readily find that $g_T(\chi_{min}^2)=0.56$, in agreement with Ref.~\cite{Benel:2019mcq}. The interpretation of the quoted uncertainties takes a different path in view of the distribution of the chi-square values in Fig.~\ref{fig:gT_chi2}. The latter is slightly shifted towards large values of $g_T$, suggesting asymmetric errors. Various replicas with low chi-square values, {\it i.e.} within $\Delta \chi^2=1$, render a large value for $g_T$. If instead we consider a $\chi^2$ distribution of $n$ degrees of freedom, the uncertainties at $68\%$ are identified by the criterion  $\Delta \chi^2=\sqrt{2 n}$, as shown on Fig.~\ref{fig:gT_chi2}, hence allowing for values of the isovector tensor charge in a broader range, {\it i.e.} $g_T\in [0,\sim 2]$. The current lattice evaluations set $g_T\sim 1$, see {\it e.g.} Ref.~\cite{Constantinou:2020hdm}.

\section{Conclusion}

In these proceedings, we have presented a follow-up analysis of the isovector tensor charge, output of a constrained fit of the valence transversity PDFs~\cite{Benel:2019mcq}. In this reference, we considered the role of the implementation of positivity bounds on the transversity PDF. This problem appeared to be related to the role of the functional form, and overall led to the conclusion that flexible parametrizations show compatibility with the lattice evaluation of the isovector tensor charge. However, the value for the up charge, $\delta u$, is systematically smaller in dihadron analyses {\it w.r.t.} lattice evaluations.

This analysis is based on semi-inclusive  dihadron data only. Dihadron-based global analyses~\cite{Radici:2018iag} include $pp$ data; global transverse-spin analyses~\cite{Cammarota:2020qcw} consider various processes and non-perturbative functions. A thorough statistical analysis of the uncertainties in those cases might shed a new light on the comparison of phenomenological results with lattice evaluations.

\section*{Acknowledgements}
The author thanks her coauthors (J. Benel and R. Ferro-Hernandez) for a fruitful collaboration, and P. Nadolsky for illuminating discussions.

\paragraph{Funding information}
The author is supported by UNAM Grant No. DGAPA-PAPIIT IA101720 and CONACyT Ciencia de Frontera 2019 No.~51244 (FORDECYT-PRONACES).



\bibliography{bib_transv.bib}

\nolinenumbers

\end{document}